\documentclass{article}

\usepackage{booktabs} 
\usepackage{authblk} 
\usepackage{graphicx} 
\usepackage{url, hyperref}
\usepackage{natbib}

\author[1]{Craig T. Russell} 
\author[2]{Jean-Marie Burel} 
\author[1]{Awais Athar} 
\author[2]{Simon Li} 
\author[1]{Ugis Sarkans} 
\author[2]{Jason Swedlow} 
\author[1]{Alvis Brazma} 
\author[1,$\ast$]{Matthew Hartley} 
\author[1,$\ast$]{Virginie Uhlmann}

\affil[1]{European Bioinformatics Institute (EMBL-EBI), European Molecular Biology Laboratory, Cambridge, UK}
\affil[2]{School of Life Sciences, University of Dundee, Dundee, UK}

\title{bia-binder: A web-native cloud compute service for the bioimage analysis community}

\begin{document}

\maketitle

\begin{abstract}
    \textbf{Summary:}
	We introduce bia-binder (BioImage Archive Binder), an open-source, cloud-architectured, and web-based coding environment tailored to bioimage analysis that is freely accessible to all researchers.
	The service generates easy-to-use Jupyter Notebook coding environments hosted on EMBL-EBI's Embassy Cloud, which provides significant computational resources.
	The bia-binder architecture is free, open-source and publicly available for deployment.
	It features fast and direct access to images in the BioImage Archive, the Image Data Resource, and the BioStudies databases.
	We believe that this service can play a role in mitigating the current inequalities in access to scientific resources across academia.
	As bia-binder produces permanent links to compiled coding environments, we foresee the service to become widely-used within the community and enable exploratory research.
	\\
	\textbf{Availability and implementation:} bia-binder is built and deployed using helmsman and helm and released under the MIT licence.
	It can be accessed at \url{binder.bioimagearchive.org} and runs on any standard web browser.
	\\
	\textbf{Contact:} \href{matthewh@ebi.ac.uk}{matthewh@ebi.ac.uk} and \href{uhlmann@ebi.ac.uk}{uhlmann@ebi.ac.uk}
\end{abstract}

\section{Introduction}
\label{sec:intro}
In recent years, the variety, complexity and volume of biological images has increased exponentially, translating into an ever-growing need for sophisticated image analysis~\citep{bagheri2022new}.
Deep learning (DL), a subset of artificial intelligence, has been instrumental in addressing the ``big data'' challenge of bioimaging and has become the \emph{de facto} standard for most of the classical bioimage analysis tasks~\citep{meijering2020bird}.
In particular, the ability of DL-based approaches to automatically recognise objects and extract relevant features from images is well underway to revolutionise how bioimages are analysed, from fundamental biology research~\citep{hallou2021deep} to biomedicine~\citep{wainberg2018deep}.
However, the potentially game-changing capabilities of DL still remains out of reach of most researchers without a strong background in computer sciences and access to high-performance computational resources.
Modern DL is indeed computationally intensive, accentuating a problem of unequal access to computing power, data and know-how.
As researchers wanting to adopt DL in their analysis require extensive computational memory as well as fast network access to image data stores, an accessibility barrier stands between the computational resources needed to address contemporary medium-to-large-scale image analysis and the now widely available methods for that purpose. A growing divide separates those who are fortunate enough to have access to computational resources and those who don't, raising concerns around fairness in access to DL-methods for bioimaging.

A number of tools have been developed towards the goal of providing easy-to-use and accessible state-of-the-art bioimage analysis methods.
In addition to well-established desktop applications such as Fiji~\citep{schindelin2012fiji}, napari~\citep{chiu2022napari}, and QuPath~\citep{bankhead2017qupath}, new web-based platforms have recently been developed.
Among them, ImJoy~\citep{ouyang2019imjoy} offers a browser-accessible computational platform that provides a user-friendly interface for developing and sharing image analysis workflows.
ZeroCostDL4Mic~\citep{chamier2020zerocostdl4mic} and its evolution DL4MicEverywhere~\citep{hidalgo2024dl4miceverywhere} adopt the different strategy of democratising advanced bioimage analysis by providing a carefully curated no-code, self-explanatory interface to interact with deep learning-based tools for various image analysis tasks, including popular methods such as multi deep-CARE~\citep{weigert2018content} for image restoration and StarDist~\citep{schmidt2018cell} for image segmentation.
These various platforms are highly valuable to address the method accessibility challenge of bioimage analysis, but are limited in their ability to scale to large datasets or to provide access to high-performance computing resources.


To address these limitations, we have implemented a cloud computing service that specifically aims to alleviate the resource availability gap in the research community.
The BioImage Archive Binder, bia-binder for short, is an open-source, cloud-based, and web-native coding environment that is freely accessible to any researcher.
The service generates Jupyter notebook coding environments that are ubiquitous in both the data science and teaching worlds, and that facilitate the quick prototyping of analysis workflows~\citep{kluyver2016jupyter}.
In the spirit of open science, environments within the service are shareable, deterministic and compliant with modern FAIR standards for data analysis~\citep{wilkinson2016fair}.


\section{bia-binder}
The bia-binder (\url{binder.bioimagearchive.org}) is a cloud-hosted web-service that allows users to directly convert public code-repositories from public hosting services such as GitHub\footnote{\url{https://github.com/}} and Zenodo\footnote{\url{https://zenodo.org/}} into interactive Jupyter notebooks through BinderHub\footnote{\url{https://binderhub.readthedocs.io/}}.
BinderHub builds deterministic environments using repo2docker\footnote{\url{https://repo2docker.readthedocs.io/}} and deploys them on Kubernetes clusters\footnote{\url{https://kubernetes.io/}} as containers running Jupyter.
The repo2docker library provides a flexible set of buildpacks that can work with several environment management tools such as conda, pip, apt, and Dockerfiles.

In addition to the base bia-binder that can be used by anyone without login, we also provide a login portal for institutional users through Elixir-AAI\footnote{\url{https://elixir-europe.org/services/aai}}, where additional resources including more RAM and CPU cores are provided (\url{login.binder.bioimagearchive.org}).
The login portal links to JupyterHub\footnote{\url{https://jupyter.org/hub}}, a gateway provider for generating online Jupyter notebooks which give access to common programming environments in notebook form, such as Python, Julia, R, along with Octave and Java-like languages through BeakerX\footnote{\url{http://beakerx.com/}}.
JupyterHub also includes Dask~\citep{gueroudji2021deisa} access, which is especially convenient for large-data tasks.
While bia-binder can automatically and conveniently convert data and code repositories into Jupyter notebook environments, the JupyterHub gives authenticated users the option to access additional resources as well as a small permanent storage area (limited to \(10\)GB for one year) for installing custom programming environments and packages alongside the Python, R, Fiji, OMERO~\citep{allan2012omero} and Julia~\citep{bezanson2017julia} environments already provided.

\begin{figure}[ht]
    \centering
    \includegraphics[width=\linewidth]{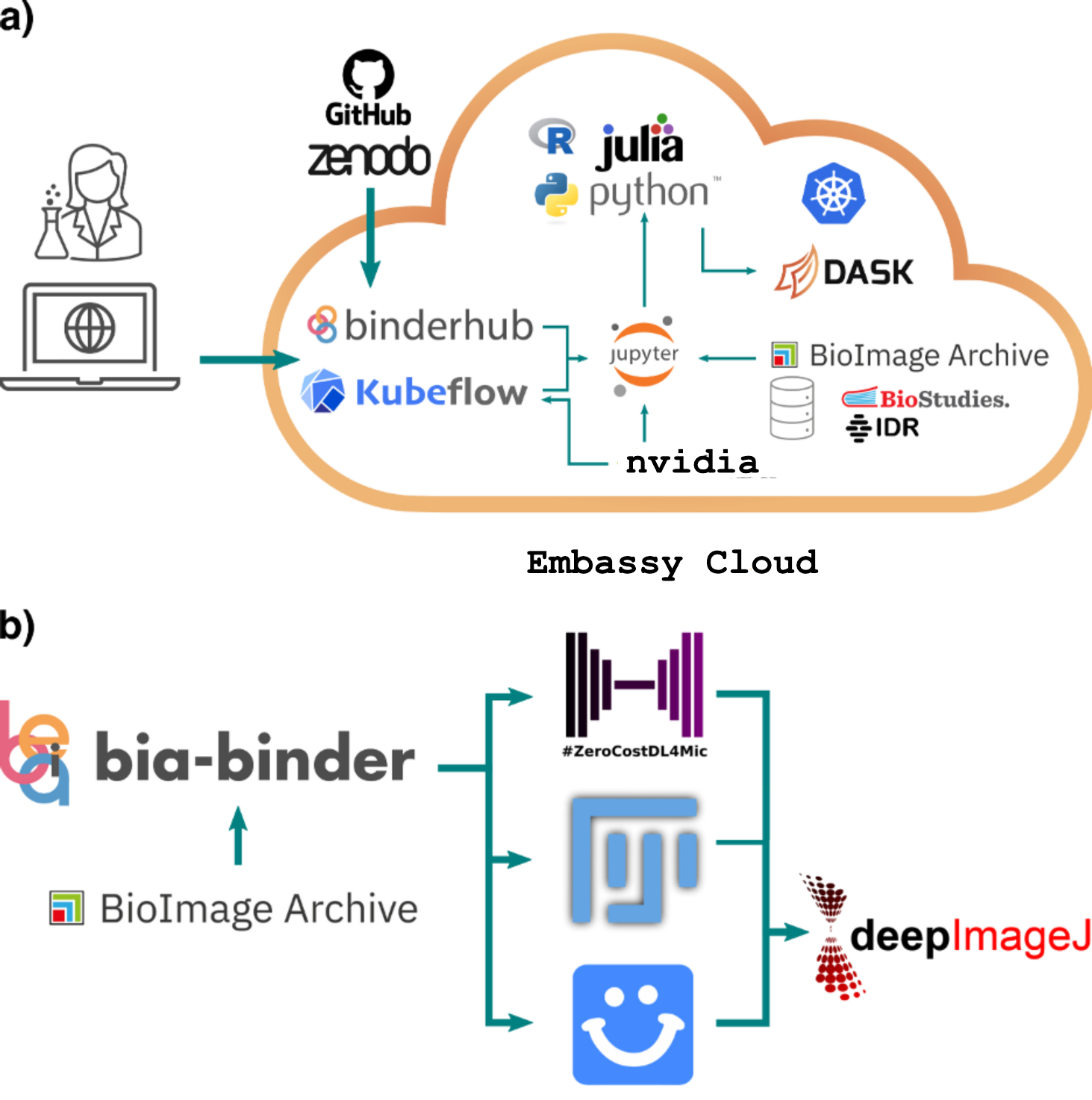}
    \caption{
    The bia-binder is deployed on Embassy Cloud hosted at EMBL-EBI. (a) It connects web-based Jupyter environments to the BioImage Archive, giving access to any Jupyter-compatible coding environment, to computational resources, and to analysis scaling tools such as Dask. (b) The bia-binder also readily integrates other available tooling ecosystems such as ZeroCostDL4Mic,  ImJoy, and Fiji.
    }
    \label{fig:fig1}
\end{figure}

In Figure~\ref{fig:fig1}, we illustrate how BinderHub and JupyterHub interact within the bia-binder environment and how bia-binder integrates with other tools in the bioimage analysis ecosystem.
As Imjoy is a web-based platform, it can be directly interacted with through Jupyter and it is also installed as a \emph{de facto} plugin for authenticated users.
Users can also interact with Fiji and the extended ImageJ ecosystem, including deepImageJ through pyImageJ~\citep{schindelin2015fiji}.


We are hosting a free, public instance of bia-binder on the EMBL-EBI Embassy Cloud\footnote{\url{https://www.embassycloud.org/}}, a service that provides significant and scalable computational resources for data analysis activities alongside EMBL-EBI’s public datasets.
As such, the Embassy Cloud is collocated with, and directly linked to, the BioImageArchive~\citep{hartley2022bioimage}, the Image Data Resource~\citep{williams2017image}, and BioStudies~\citep{sarkans2018biostudies}.
The bia-binder thus has fast and direct access to several TBs of publicly available reference image datasets, as illustrated in notebooks we provide to exemplify data access from each of these three databases.
We include several notebooks for users directly in the IDR, in the BioImage Archive through the \emph{explore in bia-binder} button, as well as on github at~\url{https://github.com/BioImage-Archive/bia-binder/tree/dev}.

Additional user-specific image data may be ingressed through standard file transfer protocols enabled here with RClone~\footnote{\url{https://rclone.org/}}, installed natively on bia-binder's login-only JupyterHub services.
Private storage areas are encrypted, no user authentication passwords are stored on the cluster, and user activities and data are neither tracked nor stored.
The bia-binder backend Kubernetes deployment is managed by Embassy Cloud and regularly maintained and updated for vulnerabilities.
For both anonymous and logged in users, active sessions are respected for up to a week before being culled and inactive sessions are given a $12$ hour lifetime to ensure that resources are freed up for other users.

The bia-binder deployment codebase is open-source and publicly available at~\url{https://github.com/BioImage-Archive/bia-binder} under the MIT license.
The bia-binder architecture is implemented to be cloud-provider agnostic and is therefore amenable to further deployment on other computational infrastructures that support Kubernetes.

We designed bia-binder to comply with FAIR data principles in that analysis pipelines are
\begin{itemize}
	\item \emph{Findable}, as BinderHub creates permanent links to data-pipelines that are stored as notebooks (.ipynb) on public repositories (access to these BinderHub images can then, in-turn, be assigned a DOI using public archiving services such as Zenodo);
	\item \emph{Accessible}, as the bia-binder service is free and publicly available;
	\item \emph{Interoperable}, as Jupyter provides native interoperability through its web-based nature and uses standard API calls and HTML, and as analysis environments are built on Docker containers, making them natively interoperable through Open Container compliance\footnote{\url{https://opencontainers.org/}};
	\item \emph{Reusable}, as documentation, example scripts and environments for interfacing with public bioimage databases such as the BioImage Archive are provided to support users in applying gold-standard analysis methods to new and archived data alike.
\end{itemize}

We primarily expect bia-binder to be interacted with as a teaching, training and data exploration tool for demonstrating state-of-the-art bioimage analysis in a convenient programming environment with attached resources.
Through its fast and native access to substantial amounts of publicly available data, bia-binder also offers an opportunity to showcase the value of bioimage data sharing and has thus the potential to contribute to a broader shift towards a more open research culture in bioimaging.

\section{Conclusions}
We have developed bia-binder to provide a convenient and capable platform for performing bioimage analysis with the wide range of cutting-edge open-source algorithms on the large corpus of publicly-available microscopy image datasets, which is growing exponentially.
We believe that bia-binder will contribute to mitigating the current inequality in access to advanced image analysis tools, especially DL-based ones, across the globe, which is chiefly driven by lack of access to computing resources.
Furthermore, as bia-binder provides permanent links to compiled coding environments through BinderHub, we foresee the service to be widely adopted by the community as a way to share interactive examples of algorithms and analyses.
Given that provision for Kubernetes cloud environments is becoming more widely available, it is possible for this service to be mirrored across multiple institutes with a single gateway federating access to the partners' deployments for analysis tasks that do not need direct access to EMBL-EBI-hosted image databases.

\section*{Acknowledgments}
CTR and MH acknowledge funding from the European Union’s Horizon Europe research and innovation programme under grant agreement 101057970 (AI4Life).
AA was supported by EOSC‐Life European programme funding from the European Union's Horizon Europe Research and Innovation Programme under grant agreement 824087.
JMB, SL, and JS acknowledges funding for the IDR project from the Wellcome Trust (ref. 212962/Z/18/Z), BBSRC (ref. BB/R015384/1) and the National Institutes of Health Common Fund 4D Nucleome Program grant UM1HG011593.
US, AB, MH and VU were supported by the European Molecular Biology Laboratory.
The authors thank Romain Laine, Guillaume Jacquemet, Ricardo Henriquez, Anna Kreshuk, Florian Jug, and Wei Ouyang for valuable comments and discussions.

\bibliographystyle{abbrvnat}
\bibliography{refs}

\end{document}